\documentclass[11pt]{article}
\pdfoutput=1

\usepackage{amssymb}
\usepackage{fullpage}
\usepackage{amsmath}
\usepackage{algorithm}
\usepackage[noend]{algpseudocode}

\newtheorem{theorem}{Theorem}
\newtheorem{lemma}[theorem]{Lemma}
\newtheorem{corollary}[theorem]{Corollary}

\newtheorem{definition}[theorem]{Definition}
\newtheorem{conjecture}[theorem]{Conjecture}

\title{The Structure of Promises in Quantum Speedups}
\author{Shalev Ben-David\thanks{Computer Science and Artificial Intelligence Lab,
				Massachusetts Institute of Technology. shalev@mit.edu}}

\begin{document}
\maketitle

\begin{abstract}
It has long been known that in the usual black-box model,
one cannot get super-polynomial quantum speedups
without some promise on the inputs.
In this paper, we examine certain types of symmetric promises,
and show that they also cannot give rise to super-polynomial quantum speedups.
We conclude that exponential quantum speedups only occur
given ``structured'' promises on the input.

Specifically, we show that there is a polynomial relationship of degree $12$ between
$D(f)$ and $Q(f)$ for any function $f$ defined on permutations
(elements of $\{0,1,\dots, M-1\}^n$ in which each alphabet element occurs exactly once).
We generalize this result to all functions $f$
defined on orbits of the symmetric group action $S_n$
(which acts on an element of $\{0,1,\dots, M-1\}^n$ by permuting its entries).
We also show that when $M$ is constant, any function $f$ defined
on a ``symmetric set'' -- one invariant under $S_n$ --
satisfies $R(f)=O(Q(f)^{12(M-1)})$.
\end{abstract}


\section{Introduction}

Quantum algorithms are generally believed to be able to solve certain
problems super-polynomially faster than any classical algorithm.
One of the most famous examples of a problem for which
a super-polynomial speedup is expected is factoring:
Shor's algorithm can be used to factor an $n$-bit integer in
$O(n^3)$ time \cite{Shor} while the best known classical algorithm
is only conjectured to achieve $e^{O(n^{1/3}\log^{2/3} n)}$ \cite{sieve}.
On the other hand, quantum computers are not believed
to be able to solve NP-complete problems in polynomial time.
It seems that quantum computers can only provide super-polynomial
speedups for certain structured problems, but not for unstructured ones.
What type of structure is required?
In this paper, we hope to help shed light on this problem.

\subsection{Query Complexity Background}

One common model for the study of quantum computation
is the \emph{black-box model} or \emph{query complexity model}.
In this model, the input is provided by adaptive queries to a black box,
each of which reveals part of the input.
The goal is to determine the value of some function $f$ on this input
(where $f$ is specified in advance), while minimizing the number of queries.
More formally, for a function $f:[M]^n\to\{0,1\}$
(where $[M]:=\{0,1,\dots, M-1\}$),
we consider an algorithm $A$ that makes adaptive queries to the entries of
$x\in [M]^n$ in order to determine $f(x)$.
The query complexity achieved by $A$ is defined to be the number
of queries required by $A$ over the worst-case choice of $x$.
The query complexity of the function $f$ is then defined to be
the minimum query complexity achieved by any algorithm $A$.

When the algorithm is deterministic, we denote the query complexity of $f$
by $D(f)$. When the algorithm is randomized (but still determines $f(x)$ with certainty),
we denote it by $R_0(f)$. If a randomized algorithm is allowed to make errors
with bounded probability (say, less than $1/3$), we denote the query complexity
by $R(f)$. Finally, if the algorithm is allowed to make quantum queries
to the input (and is also allowed to err with bounded probability), we denote
this measure by $Q(f)$.
As expected, we have the relationship
$$D(f)\geq R_0(f)\geq R(f)\geq Q(f)$$
for every function $f$. For a nice survey of query complexity
(also sometimes called decision-tree complexity), see \cite{Buhrman02}.

In the query complexity model, we can analyze the power of quantum computing
by comparing $Q(f)$ to $D(f)$ or $R(f)$. An example of an unstructured
search in this model is given by $M=2$ and $f=OR_n$, the $n$-bit OR function.
It's not hard to see that $R(f)=\Omega(n)$. On the other hand,
it has been shown that $Q(f)=\Theta(\sqrt{n})$ (the upper bound
follows from Grover search \cite{Grover}, and the lower bound was shown in \cite{bbbv}).
While the quantum query complexity is asymptotically faster than the classical
query complexity, the gap is polynomial for this unstructured problem.

An example of exponential quantum speedup can be derived from Shor's algorithm.
However, to do so, we must change the setting to allow for partial functions.
In other words, we now let $f:X\to\{0,1\}$ be defined on a set $X\subseteq [M]^n$.
We can then construct a function corresponding to a problem called period-finding.
We say $x\in [M]^n$ is periodic with period $s$ if $x_i=x_j\iff s|(i-j)$.
Let $M=\lfloor\sqrt{n}\rfloor$,
and let $X$ be the set of periodic inputs with period between
$\frac{1}{2}\sqrt{n}$ and $\sqrt{n}$. Let $f(x)$ be $0$ if the period
of $x$ is less than $\frac{3}{4}\sqrt{n}$, and $1$ otherwise.
Then any classical algorithm will require roughly $\sqrt{n}$ queries,
while Shor's algorithm requires $O(1)$.

Notice that in the period-finding problem $f$ was defined as a partial function.
We will call the set $X$ on which $f$ is defined the \emph{promise}
of the problem. In this paper, we will examine the structure of promises
that can or cannot result in super-polynomial quantum speedups.

\subsection{Previous Work}

In 1998, Beals, Buhrman, Cleve, Mosca, and de Wolf \cite{Beals01} showed
the following theorem.

\begin{theorem} \label{BinaryBeals}
If $f:\{0,1\}^n\to\{0,1\}$, then $Q(f)=\Omega(D(f)^{1/6})$.
\end{theorem}

Their result easily extends to larger alphabets:

\begin{theorem} \label{Beals}
If $f:[M]^n\to\{0,1\}$, then $Q(f)=\Omega(D(f)^{1/6})$.
\end{theorem}

This tells us that there is never a super-polynomial quantum
speedup for total functions. In fact, it is conjectured that
the relationship between $D(f)$ and $Q(f)$ is at most quadratic,
so that the OR function gives the largest gap:

\begin{conjecture} \label{ConjectureBeals}
If $f:[M]^n\to\{0,1\}$, then $Q(f)=\Omega(D(f)^{1/2})$.
\end{conjecture}

Note that these results compare quantum query complexity
to deterministic query complexity (instead of randomized).
For more information, see \cite{Buhrman02}.

Another important result was proved by Aaronson and Ambainis
in 2009 \cite{Aaronson09}. They defined a function
$f$ to be permutation-invariant if
$$f(x_1,x_2,\dots,x_n)=f(\tau(x_{\sigma(1)}),\tau(x_{\sigma(2)}),\dots,\tau(x_{\sigma(n)}))$$
for all inputs $x$ and all permutations $\sigma\in S_n$ and $\tau\in S_M$.
Here $f$ may be a partial function, but the promise set $X$ on which $f$
is defined must itself be invariant under these permutations.
As an example, if $M=2$, then $X$ might contain all binary strings
of Hamming weight in $\{1,2,n-2,n-1\}$, and $f(x)$
will depend only on the Hamming weight of $x$ (with the value of $f$
being equal on Hamming weights $k$ and $n-k$).

Aaronson and Ambainis showed the following.

\begin{theorem} \label{AaronsonAmbainis}
	If $X\subseteq [M]^n$ is permutation invariant
	and $f:X\to\{0,1\}$ is permutation invariant,
	then $Q(f)=\tilde{\Omega}(R(f)^{1/7})$.
\end{theorem}

\subsection{Our Results}

To state our results, we first require the following definition.

\begin{definition}
Given $x\in [M]^n$, the \emph{type} $\tau(x)$ of $x$
is the multiset $\{x_1,x_2,\dots,x_n\}$.

Given a type $T$ of some $x\in [M]^n$, we define $\hat{T}$ to be subset of $[M]^n$
consisting of all inputs of type $T$. Abusing notation, we will often write
$T$ instead of $\hat{T}$ to denote the set of inputs of a fixed type.
\end{definition}

For example, the type of $x=(1,1,2)$ is the multiset $\{1,1,2\}$,
so that $(1,1,2)$ and $(1,2,1)$ have the same type.
One way of thinking about this definition is as follows.
Consider the group action $S_n$ that acts on $[M]^n$ by permuting
the indices of each element $x\in[M]^n$. Then a type is simply
an orbit of this group action.

Our first result is the following theorem.


\begin{theorem} \label{type}
	If $f:T\to\{0,1\}$ is a partial function whose promise is a type, then
	$$Q(f)=\Omega\left(D(f)^{1/12}\right).$$
\end{theorem}

Note that this is a relationship between quantum query complexity
and deterministic (not randomized) query complexity.
In this sense, the result is similar to Theorem \ref{Beals},
and indeed we use some similar tools in its proof.


Our second result extends the previous theorem
from promises that are orbits of the group action
to promises that are any invariant subset
for the group action;
that is, the promise may be any ``symmetric'' set.
Unfortunately, here we are only able to prove a polynomial
relationship when $M$ is constant.

\begin{theorem} \label{symmetric}
	Let $M$ be constant.	
	If $f:X\to\{0,1\}$ is a partial function on any symmetric promise $X\in [M]$
	(that is, a set $X$ satisfying $x\in X,\sigma\in S_n\Rightarrow x_{\sigma}\in X$
	where $x_{\sigma}:=(x_{\sigma(1)},x_{\sigma(2)},\dots,x_{\sigma(n)})$)
	then
	$$Q(f)=\Omega\left(R(f)^{1/(12(M-1))}\right).$$
	In particular, when $M=2$, we have
	$$Q(f)=\Omega\left(R(f)^{1/12}\right).$$	
\end{theorem}

Unlike the previous theorem, this one only relates quantum
query complexity to randomized
(rather than deterministic)
query complexity. This is necessary; indeed, if $X$
is the set of binary strings of Hamming weight $0$ or $\lfloor n/2\rfloor$
and $f$ is defined to be $0$ on $0^n$ and $1$ elsewhere,
then $D(f)=\lfloor n/2\rfloor+1$ but $R(f)$ is constant.

Notice that this last theorem applies even to the promise $X=[M]^n$ (for constant $M$),
so it can be viewed as a generalization of Theorem \ref{BinaryBeals}
(although our polynomial relationship has higher degree,
and our generalization replaces $D(f)$ with $R(f)$).

As a final note, we remark that our results are mostly
incomparable with the Aaronson-Ambainis result (Theorem \ref{AaronsonAmbainis}).
When $M$ is constant, our Theorem \ref{symmetric}
is much more general (since it doesn't place restrictions on the function).
However, when $M$ is constant, Theorem \ref{AaronsonAmbainis}
is not very difficult in the first place;
most of the work in \cite{Aaronson09}
went towards dealing with the fact that $M$ may be large.

In the following section, we will prove Theorem \ref{type}.
Theorem \ref{symmetric} will be proven in section $3$,
and section $4$ will discuss open problems and directions for future research.

\section{Type Promises}

In this section, we show that the deterministic and quantum query complexity
measures are polynomially related when the promise is exactly
a type, proving Theorem \ref{type}.

One particular case of interest, which will motivate
a lot of our analysis, is the case where $M=n$
and $T$ is the type corresponding to the multiset $\{0,1,\dots,n-1\}$
(i.e. the case where the inputs are all permutations),
together with the function $f$ satisfying
 $f(x)=0$ iff $0$ occurs in the first $\lfloor\frac{n}{2}\rfloor$
entries of $x$. This function is sometimes called the permutation inversion
problem.

\subsection{Sensitivity, Block Sensitivity, and Certificate\\ Complexity}
We start by defining and examining sensitivity, block sensitivity,
and certificate complexity in the promise setting.
The behavior of these complexity measures is similar
on type promises to the behavior for total boolean functions
(a survey of which can be found in \cite{Buhrman02}),
with the exception that all three of these measures might be much
smaller than the deterministic query complexity
(an example of this is given by the permutation inversion problem).

We start by defining certificates.

\begin{definition}
	Let $x\in [M]^n$. A partial assignment $c$
	is an element of $([M]\cup *)^n$. $c$ is said to be
	consistent with $x$ if for all $i=1,2,\dots n$,
	either $c_i=x_i$ or $c_i=*$.
	
	Let $f:X\to\{0,1\}$ with $X\subseteq [M]^n$.
	$c$ is a $0$-certificate for $f$ if $f(x)=0$
	for all $x\in X$ consistent with $c$. Analogously,
	$c$ is a $1$-certificate for $f$ if $f(x)=1$
	for all $x\in X$ consistent with $c$.
	$c$ is a certificate if it is a $0$- or $1$-certificate.
\end{definition}

We can now define the complexity measures $C(f)$, $bs(f)$, and $s(f)$.

\begin{definition}
	Let $f:X\to\{0,1\}$ with $X\subseteq [M]$,
	and let $x\in X$. The certificate complexity $C_x(f)$ of $x$
	is the minimum size of a certificate $c$ for $f$ consistent with $x$.
	
	The certificate complexity $C(f)$ of $f$ is the maximum
	value of $C_x(f)$ out of all $x\in X$.
\end{definition}

\begin{definition}
	Let $f:X\to\{0,1\}$ with $X\subseteq [M]$,
	and let $x\in X$. The block sensitivity $bs_x(f)$ of $x$
	is the maximum size of a collection of disjoint set of indices
	$b_1,b_2,\dots\subseteq \{1,2,\dots,n\}$ (called blocks)
	such that for each block $b_i$, there is some $y\in X$
	that disagrees with $x$ only on indices in $b_i$ and for which $f(y)\neq f(x)$.
	
	The block sensitivity $bs(f)$ of $f$ is the maximum value of $bs_x(f)$
	out of all $x\in X$.
\end{definition}

Sensitivity translates somewhat less well into the promise setting.
We give the following definition for it, which makes sense primarily
when the promise is a type promise.

\begin{definition}
	Let $f:X\to\{0,1\}$ with $X\subseteq [M]$,
	and let $x\in X$. The sensitivity $s_x(f)$ of $x$
	the maximum block sensitivity of $x$ where the blocks
	all have size $2$.
	
	The sensitivity $s(f)$ of $f$ is the maximum value of $s_x(f)$
	out of all $x\in X$.
\end{definition}

Note that if we instead required the blocks to have size $1$,
then under a type promise the sensitivity of a function will always be zero,
since changing a single entry only will always break the promise.
Letting blocks have size $2$ allows two entries to be swapped,
maintaining the type promise.

We now show some relationships between $Q(f)$, $R(f)$,
$C(f)$, $bs(f)$, and $s(f)$ analogous to the ones found in
\cite{Buhrman02}.

\begin{theorem} \label{RQ}
	For all $f:X\to\{0,1\}$ with $X\subseteq [M]$,
	$R(f)=\Omega(bs(f))$ and $Q(f)=\Omega(\sqrt{bs(f)})$.
\end{theorem}

\noindent \textbf{Proof.} The proof follows by a reduction from Grover search.
Let $x$ be such that $bs(f)=bs_x(f)$, and let $b_1,b_2,\dots,b_{bs(f)}$
be disjoint sensitive blocks of $x$. Consider the input $x$ and the $bs(f)$
inputs given by changing a sensitive block of $x$. To decide the value of $f$
on such inputs, an algorithm must decide whether the input is $x$
or whether one of the blocks has been flipped; this is the setting for Grover search.
If a block was flipped, a randomized algorithm must query at
least one input from it; but this takes $\Omega(bs(f))$ queries to find.
The lower bound for $Q(f)$ follows from a simple application of
Ambainis's adversary method, as for the Grover search problem. 
\hfill$\square$

\begin{theorem}
	For all $f:X\to\{0,1\}$ with $X\subseteq [M]$,
	$s(f)\leq bs(f)\leq C(f)$.
\end{theorem}

\noindent \textbf{Proof.} $s(f)\leq bs(f)$ follows immediately from the definition.
Since a certificate must include at least one entry from every sensitive block,
we get $bs_x(f)\leq C_x(f)$ for all $x\in X$, so $bs(f)\leq C(f)$.
\hfill $\square$

\begin{theorem}
	For all $f:T\to\{0,1\}$ with $T\subseteq [M]$ a type, we have
	$C(f)\leq 3bs(f)s(f)$.
\end{theorem}

\noindent \textbf{Proof.} Let $x$ be of type $T$. Let $b_1,b_2,\dots,b_{bs_x(f)}$
be disjoint sensitive blocks of $x$, and assume each $b_i$ is minimal (under subsets).
Then $\bigcup b_i$ is a sub-certificate of $x$.

Now, we claim that the size of a sensitive block $b_i$ is at most $3s(f)$.
This gives us the desired result, because we then have a certificate of size
at most $3bs_x(f)s(f)$, which means $C(f)\leq 3bs(f)s(f)$.

Let $y\in T$ disagree with $x$ on $b_i$ with $f(y)\neq f(x)$.
Since $x$ and $y$ have the same type, the difference between them
must be a permutation on the entries of $b_i$.
In other words, there is some permutation $\sigma$ on $b_i$
such that for $j\in b_i$, we have $y_j=x_{\sigma(j)}$.

Consider the cycle decomposition $c_1c_2\dots c_k$ of $\sigma$.
Let $c_j=(a_1,a_2,\dots,a_m)$ be any such cycle.
We claim that switching $a_s$ and $a_{s+1}$ for $s\in\{1,2,\dots,m-1\}$
gives a sensitive block for $y$ of size $2$. Indeed,
if this was not a sensitive block, then block $b_i$ would not be minimal,
since $(a_s,a_{s+1})\sigma$ would be a permutation corresponding
to a smaller sensitive block (with $a_s$ removed).
Note that the number of disjoint sensitive blocks of size $2$
we can form this way is at least $\frac{|b_i|}{3}$, since
for each cycle $c_j$ we can form $\lfloor\frac{|c_j|}{2}\rfloor\geq\frac{|c_j|}{3}$
of them. Thus $s_(f)\geq \frac{1}{3}|b_i|$, as desired.
\hfill$\square$

\begin{corollary} \label{CvQ}
	Let $f:T\to\{0,1\}$ with $T\subseteq [M]$ a type. Then
	$R(f)=\Omega(C(f)^{1/2})$ and
	$Q(f)=\Omega(C(f)^{1/4})$.
\end{corollary}

\noindent \textbf{Proof.} We have
$C(f)\leq 3bs(f)s(f)\leq 3bs(f)^2$,
so $bs(f)=\Omega(\sqrt{C(f)})$. Combined with Theorem \ref{RQ},
this gives the desired result.
\hfill $\square$

\subsection{The Structure of Small Certificates}

The previous section showed a lower bound on
quantum query complexity in terms of certificate complexity
on type promises. However, this result by itself cannot
be used to relate quantum query complexity
to deterministic or randomized query complexities,
because the certificate complexity of a function
on a type promise may be much smaller than
the query complexities (an example of this is
given by the problem of inverting a permutation,
in which the certificate complexity is constant).

In this section, we prove the following technical lemma,
which will be the main tool for handling functions for
which the certificate complexity is much smaller than the
deterministic query complexity.

\begin{lemma} \label{technical}
	Let $f:T\to\{0,1\}$ with $T\subseteq [M]$ a type.
	Fix any $k\leq\frac{1}{2}\sqrt{D(f)}$.
	If $k\geq C(f)$, then there is
	\begin{itemize}
		\item a partial assignment $p$, consistent with some input of type $T$, of size at most $4k^2$, and
		\item a set of alphabet elements $S\subseteq [M]$, of size at most $4k^2$, whose
					elements each occur less than $2k$ times in $T$ outside of $p$
	\end{itemize}
	such that for any $x\in T$ which is consistent with $p$
	and any sub-certificate $c$ of $x$ of size at most $k$,
	at least one of the alphabet elements of $c-p$ is in $S$.
\end{lemma}

(Note: by $c-p$, we mean the vector $d$ with $d_i=c_i$
when $p_i=*$ and $d_i=*$ otherwise.)

Intuitively, this lemma is saying that if we restrict to inputs
consistent with $p$, then there is a small subset $S$ of the alphabet
such that an element of $S$ must exist in any small certificate.
For example, for the problem of inverting a permutation,
we can choose $p=\emptyset$, $S=\{0\}$, and $k=\lfloor n/2\rfloor-1$;
then any certificate of size less than $k$
must include the alphabet element $0$.

Our proof of this lemma is motivated by the proof
that $D(f)\leq C(f)^2$ for total boolean functions
(that proof works by repeatedly examining consistent $0$-certificates,
each of which must reveal an entry of each $1$-certificate).

\noindent \textbf{Proof of lemma.}
Fix such $T$, $f$, and $k$. The proof is based on the following algorithm,
which either generates the desired $p$ and $S$ or else computes $f(x)$ for
a given input $x$.
We will proceed by arguing that the algorithm always generates $p$ and $S$
after at most $4k^2$ queries, which must happen before it computes $f(x)$
when $x$ is the worst-case input
(as guaranteed by the requirement that $k\leq\frac{1}{2}\sqrt{D(f)}$).
The algorithm is as follows.

\begin{algorithm}
\begin{algorithmic}[1]
\State Get input $x$
\State Set $p=\emptyset$, $S=\emptyset$, $R=\emptyset$
\Loop
	\State Find any certificate $c$ (in any legal input) that
		\begin{itemize}
			\item has size at most $k$
			\item is consistent with $p$
			\item has the property that $c-p$ has no alphabet elements in $S$.
		\end{itemize}
	\State If there are no such certificates, output $p$ and $S$ and halt.
	\State Add all the alphabet elements of $c$ to $R$.
	\State Set $S$ to be the set of elements $i$ of $R$ whose multiplicity in $T$ is less than $2k$ more than the
			\indent number of times $i$ occurs in $p$.
	\State Query all domain elements of $c$ and add the results to $p$.
	\State If $p$ is a $0$-certificate, output "$f(x)=0$" and halt; if it's a $1$-certificate, output "$f(x)=1$" 
			\indent and halt.
\EndLoop
\end{algorithmic}
\end{algorithm}

We claim that this algorithm will go through the loop at most $4k$ times.
Indeed, each iteration through the loop selects a certificate.
A $0$-certificate must conflict with all $1$-certificates, and vice versa.
There are two ways for certificates to conflict:
either they disagree on the value of an entry, or else there is some
alphabet element $i$ that they claim to find in different places
(and in addition, there must be few unrevealed instances of $i$ in $x$).

This motivates the following definition: for a certificate $c$,
let $h_{p,S}(c)$ be $|c-p|+|\operatorname{alphabet}(c)-S|$ if $c$ is consistent with $p$,
and zero otherwise (here $|c-p|$ denotes the number of non-$*$ entries in the partial assignment $c-p$,
and $\operatorname{alphabet}(c)$ denotes the set of alphabet elements occurring in $c$).
Note that at the beginning of the algorithm,
$h_{p,S}(c)\leq 2|c|\leq 2k$ for all certificates $c$ of size at most $k$.
Now, whenever the algorithm considers a $0$-certificate $c_0$, the value of
$h_{p,S}(c_1)$ decreases for all $1$-certificates $c_1$.
This is because either $c_0$ and $c_1$ conflict on an input, in which
case an input is revealed, decreasing $|c_1-p|$ (or contradicting $c_1$),
or else $c_0$ and $c_1$ both include a range element $i$
which has less than $2k$ occurrences left to be revealed according to $T$
(if it had at least $2k$ unrevealed occurrences,
it wouldn't be the source of a conflict between $c_0$ and $c_1$,
since they each have size at most $k$).
In the latter case, $i$ is added to $S$, which decreases $|\operatorname{range}(c_1)-S|$.

We have shown that each iteration of the algorithm
decreases $h_{p,S}(c)$ either for all $0$-certificates or
for all $1$-certificates (of size at most $k$).
This means that unless the loop is terminated,
one of the two values will reach $0$ in less than
$4k$ iterations. We claim this cannot happen, implying
the loop terminates in less than $4k$ iterations.

Without loss of generality, suppose by contradiction that
$h_{p,S}(c)$ reaches $0$ for all $0$-certificates.
This means $p$ is either a certificate - in which case the value of $f(x)$
was determined, which is a contradiction - or else $p$ is not a certificate, and
conflicts with all $0$-certificates of size at most $k$.
In the latter case, there is some input $y$ consistent with $p$
such that $f(y)=0$, and this input cannot have $0$-sub-certificates
of size at most $k$. Thus $C(f)>k$, contradicting the assumption
in the lemma.

This shows the loop always terminates in less than $4k$ iterations,
which means it cannot calculate $f(x)$, and must instead
output $p$ and $S$. This gives the desired result,
since all certificates of size at most $k$ that are consistent with $p$
have the property that $c-p$ has a range element in $S$.
\hfill$\square$

Note that if we restrict to inputs consistent with $p$,
then the lemma asserts that finding a small certificate
requires finding an element of $S$.
This gives us the following corollary:

\begin{corollary} \label{R0}
		If $f:T\to\{0,1\}$ with $T\subseteq [M]$ a type, then we have
	$$R_0(f)=\Omega(\min(D(f)^{1/2},n^{1/4}))=\Omega(D(f)^{1/4}).$$
	When $M=n$ and $T$ is the type of permutations, we have
	$$R_0(f)=\Omega(\min(D(f)^{1/2},n^{1/3}))=\Omega(D(f)^{1/3}).$$
\end{corollary}

\noindent \textbf{Proof.}
Fix $T$ and $f$, and let $k=\lfloor\min(\frac{1}{2}\sqrt{D(f)},\frac{1}{4}n^{1/4})\rfloor-1$
(in the case of permutations, let $k=\lfloor\min(\frac{1}{2}\sqrt{D(f)},\frac{1}{4}n^{1/3})\rfloor-1$).
Since a zero-error randomized algorithm must find a certificate,
if $k<C(f)$, the desired result follows.
It remains to treat the case where $k\geq C(f)$.

In this case, let $p$ and $S$ be as in the lemma.
We restrict to inputs consistent with $p$.
Any zero-error randomized algorithm must find a certificate on such inputs.
Suppose by contradiction that algorithm $A$ has the property that on any such
input, it requires at most $k$ queries with probability at least $\frac{1}{2}$.

In order to query at most $k$ times, $A$ would need to
find a certificate of size at most $k$. But this means that on all inputs $x$, $A$
finds an element of $S$ in $x$ outside $p$ with probability at least $\frac{1}{2}$.
However, there are at most $2k|S|=8k^3$ such elements in the entries of $x$ outside $p$
(in the case of permutations, at most $|S|=4k^2$ such elements),
and the size of the domain is $n-|p|\geq n-4k^2\geq \frac{n}{2}$.
If $x$ is generated by fixing $p$ and permuting the remaining
entries randomly, the chance of a query finding an element of $S$ is
thus at most $\frac{16k^3}{n}$, so by the union bound,
the chance of finding such an element after $k$ queries is at most
$\frac{16k^4}{n}$ (in the case of permutations, this becomes
$\frac{8k^3}{n}$). Choosing $k<\frac{1}{2}n^{1/4}$
(or $k<\frac{1}{2}n^{1/3}$ in the case of permutations)
gives the desired contradiction.

We conclude any zero-error randomized algorithm must make
at least $\Omega(k)$ queries, which gives the desired result.
\hfill$\square$

\subsection{Lower bounds on R(f) and Q(f)}

We now put everything together to prove lower bounds on $R(f)$
and $Q(f)$ in terms of $D(f)$, proving Theorem \ref{type}.

\begin{theorem} \label{strongtype}
	For any $f:T\to\{0,1\}$ with $T\subseteq [M]^n$ a type, we have
	$$R(f)=\Omega(D(f)^{1/6})$$
	and
	$$Q(f)=\Omega(D(f)^{1/12}).$$
\end{theorem}

(Note that unlike Corollary \ref{R0}, we do not get an improvement here
for the special case of permutations.)

The proof of this theorem will require a version of Ambainis's
adversary method \cite{Ambainis}, which we quote here for convenience.

\begin{theorem} \label{adversary}
	Let $f:X\to \{0,1\}$ with $X\subseteq [M]^n$.
	Let $A,B\subseteq X$ be such that $f(a)=0$ for all $a\in A$
	and $f(b)=1$ for all $b\in B$. Let $R\subseteq A\times B$ be such that
	\begin{enumerate}
		\item For each $a\in A$, there exist at least $m$ different $b\in B$ such that $(a,b)\in R$.
		\item For each $b\in B$, there exist at least $m'$ different $a\in A$ such that $(a,b)\in R$.
	\end{enumerate}
	Let $l_{a,i}$ be the number of $b\in B$ such that $(a,b)\in R$ and $a_i\neq b_i$.
	Let $l_{b,i}$ be the number of $a\in A$ such that $(a,b)\in R$ and $a_i\neq b_i$.
	Let $l_{max}$ be the maximum of $l_{a,i}l_{b,i}$ over all $(a,b)\in R$ and $i\in\{1,2,\dots, n\}$
	such that $a_i\neq b_i$. Then $Q(f)=\Omega\left(\sqrt{\frac{mm'}{l_{max}}}\right)$.
\end{theorem}

\noindent\textbf{Proof of Theorem \ref{strongtype}.}
Apply Lemma \ref{technical} with $k=\frac{1}{4}\sqrt{D(f)}$.
If $C(f)>k$, then we're done by Corollary \ref{CvQ}.
Otherwise, we get $p$ and $S$ from the lemma,
with all certificates of size at most $k$ that are consistent
with $p$ having range elements in $S$.

We use Ambainis's adversary method to get a lower
bound for $Q(f)$, which will look very similar to
the lower bound for permutation inversion found in \cite{Ambainis}.
In order to construct the sets for the adversary method,
we use the following procedure.

\begin{algorithm}
\begin{algorithmic}[1]
\State Set $r=\emptyset$.
\While{$|r|\leq k$}
	\State Pick any $0$-certificate $c$ consistent with $p$ and $r$
			of size at most $C(f)$.
	\State Add the entries of $c$ to $r$, but replace any alphabet element in $S$ with an arbitrary
			\indent alphabet element in $[M]-S$.
	\State If $|r|>k$, stop. Otherwise, repeat steps $3$ and $4$ for a $1$-certificate.
\EndWhile
\end{algorithmic}
\end{algorithm}

Note that $r$ never contains alphabet elements in $S$.
Thus as long as $|r|\leq k$, the partial assignment $p\cup r$
cannot be a certificate, by the lemma. This means the selection
of certificates in steps $3$ and $5$ cannot fail.
Each iteration of the loop increases $|r|$ by at most
$2C(f)$, so this loop repeats at least
$\frac{k}{2C(f)}-1$ times.

Consider the subsets of $r$ that were added
by the selection of $0$-certificates. Let them be
$c^{(0)}_1,c^{(0)}_2,\dots,c^{(0)}_{\alpha}$,
with $\alpha\geq \frac{k}{2C(f)}-1$.
Similarly, let the subsets of $r$ that were added by $1$-certificates
be $c^{(1)}_1,c^{(1)}_2,\dots,c^{(1)}_{\alpha}$.
Note that if some of the alphabet elements in $c^{(j)}_i$
were replaced by some elements from $S$,
we would get a $j$-certificate.
We use this fact to construct the sets for the adversary method.

Let $A$ be the multiset of alphabet elements of the selected certificates
that are in $S$. Since the total size of the certificates selected
is $|r|\leq 2k$, we have $|A|\leq 2k$.

To each $c^{(j)}_i$, we add an arbitrary block of
$|A|$ entries outside $p$ and $r$ with alphabet elements outside $A$.
To be able to do this, we require that
$2|A|\alpha\leq n-|r|-|p|-2k|A|$ (the $2k|A|$ term appears
because each alphabet element in $A$ may occur up to $2k$ times).
Since $|r|,|A|\leq 2k$ and $|p|\leq 4k^2$, it suffices to have
$\alpha\leq\frac{n-10k^2}{4k}$.
Since $k$ satisfies $k\leq\frac{1}{4}\sqrt{n}$, the right hand side is within
a constant factor of $\frac{n}{k}$.
We restrict $\alpha$ to this value if it was larger than it.

Now we can place all the alphabet elements of $S$
inside any $c^{(j)}_i$ in a way that restores the $j$-certificate.
We can thus generate $2\alpha$ inputs, $\alpha$ of which
have value $0$ and $\alpha$ of which have value $1$,
such that the only difference between the inputs is
which of the $2\alpha$ disjoint bins have the alphabet elements of $S$.
This is essentially a version of permutation inversion.

It's clear that a classical randomized algorithm must
make $\Omega(\alpha)$ queries, since it must find
the bin containing the alphabet elements of $S$.
For the quantum lower bound, we use Theorem \ref{adversary}.
Let $A$ be the set of indices in which the elements of $S$
were placed for a $0$-certificate bin, and let $B$
be the set of indices in which the elements of $S$
were placed for a $1$-certificate bin.
Our relation $R$ will simply be $A\times B$.
Then each element of $A$ has $\alpha$ neighbors in $B$,
and vice versa. However, for each domain entry $q$ and $(a,b)\in R$,
we have $l_{a,q}=1$ or $l_{b,q}=1$, so $l_{a,q}l_{b,q}\leq\alpha$.
Thus we get a quantum lower bound of $\Omega(\sqrt{\alpha})$.

Finally, to complete the proof, we note that
$\alpha=\Omega(\min(\frac{n}{k},\frac{k}{C(f)}))=\Omega(\frac{k}{C(f)})$ (since $n>k^2$),
so that, combining with corollary,
$R(f)=\Omega(\beta)$
and $Q(f)=\Omega(\sqrt{\beta})$
with
$\beta=\max(\sqrt{C(f)},\frac{k}{C(f)})$.
Note that this satisfies $\beta=\Omega(k^{1/3})$.
Picking $k=\frac{1}{4}\sqrt{D(f)}$
gives $\beta=\Omega(D(f)^{1/6})$,
as desired.
\hfill$\square$

\section{Symmetric Promises with Small Range}

In this section, we show a polynomial relationship
between $Q(f)$ and $R(f)$ for any function on a symmetric
promise whose range is constant, proving Theorem \ref{symmetric}.
We will use the term symmetric to refer to invariance under
permutation of the indices of the inputs.

\subsection{The case of Symmetric Functions}
We start by dealing with the case where the function $f$
is itself symmetric. We prove the following theorem.

\begin{theorem} \label{sym}
	Let $M$ be a constant, let $X\subseteq [M]^n$ be symmetric,
	 and let $f:X\to\{0,1\}$ be a symmetric function
	 (so that $f(x)$ depends only on the type of $x$.)
	 Then $$Q(f)=\Omega\left(\frac{R(f)^{1/8}}{M\log^{1/8} M}\right).$$

\end{theorem}

In order to prove this theorem, we relate $Q(f)$ and $R(f)$
to a new complexity measure $g(f)$, which we now define.

\begin{definition}
	If $M$ is a constant and $T_1,T_2$ are types with range $M$,
	then the distance $d(T_1,T_2)$ between $T_1$ and $T_2$
	is the maximum over all $i\in[M]$ of the difference between
	the multiplicity of $i$ in $T_1$ and the multiplicity of $i$ in $T_2$.
	
	If $f:X\to\{0,1\}$ is a symmetric function with a symmetric promise $X\subseteq [M]^n$,
	define $d(f)$ to be the minimum value of $d(T_1,T_2)$
	for types $T_1,T_2\subseteq X$ that have different value under $f$.
	Define $g(f):=\frac{n}{d(f)}$.
\end{definition}

We proceed to prove lemmas relating $g(f)$ to $R(f)$ and $Q(f)$ to $g(f)$.

\begin{lemma}\label{sampling}
	For any $x\in [M]$, $O(\frac{n^2\log M}{d^2})$
	queries suffice to find a type $T$ such that
	$d(T,\tau(x))<d$ with probability at least $\frac{2}{3}$
	(where $\tau(x)$ denotes the type of $x$).
	Hence, if $f:X\to\{0,1\}$ is symmetric, then
	$R(f)=O(g(f)^2\log M)$.
\end{lemma}

\noindent \textbf{Proof.} We describe a classical randomized algorithm
for estimating the type of input $x$. The algorithm is simply the
basic sampling procedure that queries
random entries of $x$ and keeps track of the number $r_i$
of times each range element $i$ was observed.
The type $T$ is then formed by $T(i)=\frac{r_i}{\sum_{i\in[M]}r_i}n$.

Let the type of $x$ be $\tau(x)=(t_1,t_2,\dots,t_M)$,
so that the multiplicity of range element $i$ in $\tau(x)$ is $t_i$.

A version of the Chernoff bound states that if we have
$k\geq\frac{3}{\epsilon^2}\ln\frac{2}{\delta}$
samples estimating the proportion $p$ of the population
with some property, the proportion of the sample with
that property is in $(p-\epsilon,p+\epsilon)$
with probability at least $1-\delta$.
Setting $\epsilon=\frac{d}{n}$ and $\delta=1-\frac{1}{3M}$,
we see that $O(\frac{n^2\log(M)}{d^2})$ samples
suffice for $\frac{T(i)}{n}$ to be within $\frac{d}{n}$
of $\frac{t_i}{n}$ with probability at least $1-\frac{1}{3M}$.
In other words, we have $|T(i)-t_i|<d$ with probability $1-\frac{1}{3M}$
for each $i$.

The union bound then gives us $|T(i)-t_i|<d$ for all $i$
with probability at least $\frac{2}{3}$. This shows that
$d(T,\tau(x))<d$, as desired.

To compute $f(x)$ for symmetric $f$, a randomized algorithm
can estimate the type of $x$ to within $\frac{d(f)}{2}$, and
then just output the value of $f$ on any input of type
within $\frac{d(f)}{2}$ of the estimated type $T$.
Since $g(f)=\frac{n}{d(f)}$, we get $R(f)=O(g(f)^2\log M)$.
\hfill$\square$

\begin{lemma}\label{gvQ}
	If $f:X\to\{0,1\}$ is symmetric
	with a symmetric promise $X\subseteq [M]^n$, then
	$$Q(f)=\Omega\left(\frac{g(f)^{1/4}}{M}\right).$$
\end{lemma}

\noindent \textbf{Proof.}
Let $S$ and $T$ be types with distance $d(f)$
such that if $x$ has type $S$ and $y$ has type $T$
then $f(x)\neq f(y)$. We claim that a quantum algorithm
cannot distinguish between these types in less than the
desired number of queries.

We proceed by a hybrid argument. We form a sequence
of types $\{S_i\}_{i=0}^k$ with $k\leq M$ such that $S_0=S$, $S_k=T$,
and for all $i=0,1,\dots,k-1$, the types $S_i$ and $S_{i+1}$
differ in the multiplicity of at most $2$ range elements and
have distance at most $d(f)$.

We do this as follows. Set $S_0=S$. Let $A$ be the set
of range elements on whose multiplicities the current $S_i$
agrees with $T$; at the beginning, $A$ is the set of range elements
on which $S$ and $T$ have the same multiplicity, which may be empty.
To construct $S_{i+1}$ given $S_i$,
we simply pick a range element $r$ for which $S_i$ has a larger multiplicity
than $T$ and a range element $r'$ for which $S_i$ has a smaller multiplicity
than $T$. We then set $S_{i+1}$ to have the same multiplicities as $S_i$,
except that the multiplicity of $r$ is reduced to that in $T$ and the
multiplicity of $r'$ is increased to make up the difference.
Note that the multiplicity of $r$ is then equal in $S_i$ and $T$, so $r$
gets added to $A$. Moreover, note that $d(S_i,S_{i+1})\leq d(S_i,T)$,
and also $d(S_{i+1},T)\leq d(S_i,T)$. Since this is true for all $i$, it follows that
$d(S_i,S_{i+1})\leq d(S,T)=d(f)$.

Since a range element gets added to $A$ each time and these elements
are never removed, this procedure is terminated with $S_k=T$ after at most $M$ steps.
Thus $k\leq M$. In addition, consecutive types differ in the multiplicities of $2$
elements and have distance at most $d(f)$.

We now give a lower bound on the quantum
query complexity of distinguishing $S_i$ from $S_{i+1}$.
Without loss of generality, let the range elements for which
$S_i$ and $S_{i+1}$ differ be $0$ and $1$,
with $0$ having a smaller multiplicity in $S_i$.
Let $a$ be the multiplicity of $0$ in $S_i$, and let $b$
be the multiplicity of $1$ in $S_i$, with $0<b-a\leq d(f)$.
Let $c$ and $d$ be the multiplicities of $0$ and $1$ in $S_{i+1}$,
respectively. Then $c+d=a+b$. Let $e=a+b=c+d$.

We prove two lower bounds using Ambainis's adversary
method, corresponding to $e$ being either large or small.
For the small case, consider an input $x$ of type $S_i$
split into $2\alpha=\lfloor\frac{n}{e}\rfloor$
blocks $B_1,B_2,\dots,B_{2\alpha}$ of size $e$ each, such that
all the $0$ and $1$ elements lie in block $B_1$.
To change the input from type $S_i$ to $S_{i+1}$,
we must simply change the first block. Also,
note that rearranging the blocks does not change the type.
Let $X$ be the set of inputs given by rearranging the blocks of $x$
so that the block $B_1$ ends up in the first $\alpha$ blocks, and let
$Y$ be the set of inputs given by replacing $B_1$ to get type $S_{i+1}$
and then rearranging the blocks so that $B_1$ ends up
in the last $\alpha$ blocks. We now have a reduction from
the problem of inverting a permutation, so using Ambainis's adversary
method, we get a lower bound of $\Omega(\sqrt{\alpha})=\Omega(\sqrt{\frac{n}{e}})$.

For the case when $e$ is large, we restrict to inputs in which
all elements are fixed except for those that have value $0$ or $1$.
The lower bound of $\Omega(\sqrt{\frac{e}{d(f)}})$ then
follows from Lemma 22 in \cite{Aaronson09}.

If $e\leq \sqrt{nd(f)}$, the former bound gives
a lower bound of $\Omega((\frac{n}{d(f)})^{1/4})$
for distinguishing $S_i$ from $S_{i+1}$ by quantum queries.
If $e\geq\sqrt{nd(f)}$, the latter bound gives the same.
Thus we have a lower bound of $\Omega(g(f)^{1/4})$
in all cases.

Finally, note that if a quantum algorithm could compute $f(x)$
in $Q(f)$ queries, then for some $i$ it could distinguish
$S_i$ from $S_{i+1}$ with probability $\Omega(\frac{1}{M})$.
This means we could use $MQ(f)$ queries to distinguish $S_i$ from $S_{i+1}$
with constant probability, so $Q(f)=\Omega(\frac{1}{M}g(f)^{1/4})$.
\hfill$\square$

These two lemmas combine to prove Theorem \ref{sym},
which can be restated as the following corollary.

\begin{corollary}
	For symmetric $f$ on alphabet of size $M$, we have
	$$R(f)=O(Q(f)^8 M^8\log M).$$
\end{corollary}

\subsection{The General Case}

In this section, we prove Theorem \ref{symmetric}.
The proof proceeds by describing a classical algorithm
that doesn't use too many more queries than the best quantum algorithm.
An interesting observation is that this classical algorithm
is mostly deterministic, and uses only
$O(Q(f)^8 M^8 \log M)$ randomized queries at the beginning
(in order to estimate the type of the input).

\noindent \textbf{Proof.} Let $f$ be a function. We describe a classical
algorithm for computing $f$ on an input $x$, and argue that
a quantum algorithm cannot do much better.

As a first step, the algorithm will estimate the type of $x$
using $O(Q(f)^8 M^8\log M)$ queries.
By lemma \ref{sampling},
this will provide a type $T$ such that $d(T,\tau(x))<\frac{n}{cM^4Q(f)^4}$
with high probability, where we choose the constant
$c$ to be larger than twice the asymptotic constant in Lemma \ref{gvQ}.
We restrict our attention to types that are within $\frac{n}{cM^4Q(f)^4}$ of $T$.

For this proof, we will often deal with certificates $c$ for $f$
that only work on inputs of some specific type $S$; that is,
all inputs $x\in X$ of type $S$ that are consistent with $c$
have the same value under $f$.
We will say $c$ is a certificate for the type $S$.

Now, notice that if we fix a type $S$ and assume that $x$
has this type, then there is a deterministic algorithm that
determines the value of $f(x)$ in at most $\alpha$ steps,
where $\alpha=O(Q(f)^{12})$. Since this is a deterministic
algorithm, it must find a certificate of size at most $\alpha$
for the type $S$. The only other possibility is that
the deterministic algorithm finds a partial assignment
that contradicts the type $S$, in which case it cannot proceed.
Running this deterministic algorithm on type $S$ will be called
\emph{examining} $S$.

Note further that we can never find a $0$-certificate $c_0$
for some type $S_0$ and a $1$-certificate $c_1$ for some
other type $S_1$ without the certificates contradicting either type.
This is because if we found such certificates, then
fixing those entries and shuffling the rest according
to either $S_0$ or $S_1$ will give two types $S_0-(c_0\cup c_1)$
and $S_1-(c_0\cup c_1)$ on inputs of size $n-|c_0\cup c_1|$
with distance at most $\frac{n}{cM^4Q(f)^4}$
that the quantum algorithm must distinguish between.
Since $n>4\alpha$ (or else $R(f)=O(n)=O(Q(f)^{12})$),
we have $n-|c_0\cup c_1|>\frac{n}{2}$,
and $(\frac{n}{2})/(\frac{n}{cM^4Q(f)^4})=\frac{cM^4Q(f)^4}{2}$;
then Lemma \ref{gvQ} together with the choice of $c$
imply that a quantum algorithm takes more than $Q(f)$ queries
to distinguish these types, giving a contradiction.

For a type $S$, we now define $v(S)\in [2\alpha+1]^M$
to be the vector with $v(S)_i=\min(S(i),2\alpha)$ for all $i$,
where $S(i)$ is the multiplicity of $i$ in the type $S$.
If an input has type $S$, we call $v(S)$ the simplified type
of the input. We consider the partial order on simplified types
given by $v(S)\geq v(R)$ if and only if $v(S)_i\geq v(R)_i$ for all $i=1,2,\dots M$.
We say a simplified type $v(S)$ is maximal if it is maximal in this partial order.

The algorithm proceeds by finding the set of maximal simplified types,
and selecting a representative type $S$ for each maximal simplified type $v$
so that $v(S)=v$. Let the types selected this way be $S_1,S_2,\dots,S_{\beta}$.
For each $S_i$, we then run the deterministic algorithm that uses $\alpha$ queries
assuming type $S_i$. Let $c_i$ be the set of queries made by this algorithm for type $S_i$.
Note that the total number of queries made this way is at most $\alpha\beta$.

For each $S_i$, the partial assignment $c_i$ is either a certificate for $S_i$
or a disproof of the type $S_i$. Consider the pairwise unions $c_i\cup c_j$.
We restrict our attention to the types $S_i$ that are consistent with $c_i\cup c_j$
for all $j$. We claim that there is at least one such type.
Indeed, if $T$ is the true type of the input,
then $v\geq v(T)$ for some maximal simplified type $v$, and $v(S_k)=v$ for some $k$.
Then $S_k$ cannot be disproven in $2\alpha$ queries, as that would disprove $v$
and therefore $v(T)$ as well.

Now, let $S_i$ and $S_j$ be any two types remaining.
Then they are both consistent with $c_i\cup c_j$.
As we saw earlier, we cannot have $c_i$ be a $0$-certificate for $S_i$
and $c_j$ be a $1$-certificate for $S_j$ (or vice versa); the certificates
$c_i$ and $c_j$ must agree. We conclude that the certificates
$c_i$ for the remaining types are either all $0$-certificates (for their respective types)
or all $1$-certificates. Our algorithm will then output $0$ in the former case
and $1$ in the latter.

To see that the algorithm is correct, recall that
$S_k$ is one of the remaining types, with $v(S_k)=v\geq v(T)$.
Without loss of generality, suppose the algorithm output $0$, so that $c_k$
is a $0$-certificate. Suppose by contradiction that $f(x)=1$
for the our input. Let $c$ be a $1$-certificate consistent with $x$ of size at most $\alpha$.
Then $c$ is a $1$-certificate for the type $T$.
Now, $c\cup c_k$ cannot disprove $v(T)$ (since it has size at most $2\alpha$),
so $c\cup c_k$ cannot disprove $T$. Since $c\cup c_k$ cannot disprove $v(T)$,
it also cannot disprove $v$, so it cannot disprove $S_k$.
This means $T$ and $S_k$ are not disproven by their $0$- and $1$-certificates,
which we've shown is a contradiction. Thus if the algorithm outputs $0$, we must have $f(0)$ as well.

The total number of queries required is
$O(Q(f)^8 M^8\log M)+\alpha\beta$,
where $\alpha=O(Q(f)^{12})$. We must estimate $\beta$,
the number of maximal simplified types. This is at most the number of
maximal elements in $[2\alpha+1]^M$ in our partial order.
We can show by induction that this is at most $(2\alpha+1)^{M-1}$:
in the base case of $M=1$, the value is $1$, and when $M$ increases
by $1$ the number of maximal elements can increase by at most a factor of $(2\alpha+1)$.
This gives a final bound of $O(Q(f)^{12M})$ on the number of queries when $M$ is constant.

To reduce this to $O(Q(f)^{12(M-1)})$, we note that
some alphabet element $a$ must occur at least $n/M$ times in $T$,
by the pigeonhole principle. We could then use $O(M\alpha)$ queries
to find $2\alpha$ instances of $a$ with high probability.
Then each simplified type $v$ will have $v_a=2\alpha$,
so the simplified types are effectively elements of
$[2\alpha+1]^{M-1}$ instead of $[2\alpha+1]^M$.
This decreases $\beta$ to $(2\alpha+1)^{M-2}$,
so the total number of queries decreases to $O(Q(f)^{12(M-1)})$.

\hfill$\square$

\section{Conclusion}

In this paper, we have shown that certain types of
promises do not suffice for quantum speedups.
These promises are highly symmetric; we could say that
they lack structure that a quantum algorithm could exploit.

One natural question is whether we could expand these
results to symmetric promises with large alphabets.
Such a result would generalize the result of Aaronson and
Ambainis (Theorem \ref{AaronsonAmbainis}).
Proving such a theorem seems tricky; in fact,
even the case of symmetric functions with symmetric promises
was left as a conjecture in Aaronson and Ambainis \cite{Aaronson09}.

One observation is that Aaronson and Ambainis managed to overcome
the difficulties posed by a large alphabet by requiring
a symmetry on the alphabet elements as well.
Perhaps expanding that result to promises that satisfy
both symmetries would be more tractable.

One of the strongest possible versions of these results could be as follows.

\begin{conjecture}
	Let $f:X\to\{0,1\}$ with $X\subseteq [M]^n$ symmetric.
	Then $Q(f)=\Omega(R(f)^{1/2})$.
\end{conjecture}

This conjecture was pointed out to me by Aaronson (personal communication).
It says that a Grover speedup is the best a quantum algorithm
can achieve on a symmetric promise.
There does not seem to be a known counterexample to this conjecture.

Even more generally, we can ask the question of what kinds
of symmetries suffice for exponential quantum speedups.
In other words, let $G$ be a group action which acts on $[M]^n$
by permuting the indices of each element $x\in [M]^n$.
For which groups $G$ can a $G$-invariant promise
yield a super-polynomial quantum speedup?
Shor's algorithm demonstrates such a speedup
when $G$ is a cyclic group. The results in this paper
suggest that there may not be a speedup when $G$
is the symmetric group. It would be interesting
to analyze this question for other groups $G$.

\bibliographystyle{plain}


%
%
%
%
%

\end{document}